\documentclass[useAMS,usenatbib]{mn2e}
\usepackage{aas_macros,graphicx,multirow,amsmath}
\usepackage[]{color}
\usepackage{subfigure}
\usepackage{latexsym}
\usepackage{hyperref}
\usepackage{times} 

\graphicspath{{plots/}}

\def\be{\begin{equation}}
\def\ee{\end{equation}}
\def\ba{\begin{eqnarray}}
\def\ea{\end{eqnarray}}

\begin{document}

\def\LaTeX{L\kern-.36em\raise.3ex\hbox{a}\kern-.15em
    T\kern-.1667em\lower.7ex\hbox{E}\kern-.125emX}

\def\xmm {\emph{XMM-Newton}}
\def\cxo {\emph{Chandra}}
\def\swift {\emph{Swift}}
\def\sax {\emph{BeppoSAX}}
\def\rxte {\emph{RXTE}}
\def\rst {\emph{ROSAT}}
\def\flux {\mbox{erg cm$^{-2}$ s$^{-1}$}}
\def\lum {\mbox{erg s$^{-1}$}}
\def\v {\upsilon}

%\title[Jets and gamma-ray emission from isolated accretion black holes]{Jets and gamma-ray emission
%from isolated accretion black holes}
\title[GMM and population synthesis of radio pulsars]{Gaussian mixture
  models and the population synthesis of radio pulsars}

\author[A. Igoshev and S. Popov]{A.P. Igoshev$^{1}$\thanks{E-mail: ignotur@gmail.com} 
and 
S.B.~Popov$^{2}$\thanks{E-mail: sergepolar@gmail.com}\\
\smallskip\\
$^1$ Sobolev Institute of Astronomy, Saint Petersburg State University, Universitetsky prospekt 28, 198504, Saint Petersburg, Russia\\
$^2$ Sternberg Astronomical Institute, Lomonosov Moscow State University, 
Universitetsky prospekt 13, 119991, Moscow, Russia  }

%\date{Accepted \ldots. Received \ldots; in \\
%original form \ldots} \pagerange{\pageref{firstpage}--\pageref{lastpage}} \pubyear{2011}

\date{}

%\begin{document}

%\label{firstpage}
\maketitle
\begin{abstract}
Recently, Lee et al. used Gaussian mixture models (GMM) to study the
radio pulsar population. In the distribution of normal pulsars in the
$P$~--~$\dot P$ plane, they found four clusters. We develop this
approach further and apply it to different synthetic pulsar
populations in order to determine whether the method can effectively
select groups of sources that are physically different.  We check
several combinations of initial conditions as well as the models of
pulsar evolution and the selection effects. We find that in the case
of rapidly evolving objects, the GMM is oversensitive to parameter
variations and does not produce stable results. We conclude that the
method does not help much to identify the sub-populations with
different initial parameters or/and evolutionary paths. For the same
reason, the GMM does not discriminate effectively between theoretical
population synthesis models of normal radio pulsars.
\end{abstract}

\section{Introduction}
\label{intro}

While radio pulsars are known for more than 45 years \citep{bell1968},
many aspects of their initial properties and the evolution are not
well understood. The evolution of a pulsar is a long process and
cannot be studied directly.  Fortunately, about 2000 pulsars at
different stages of their evolution are currently being observed
(\citealt{atnf}). Therefore, since the early seventies, different
methods of statistical analysis of the radio pulsar population and its
evolution have been proposed (see \citealt{vive1981,vran2011} and
references therein).  The latest attempt was made by \cite{lee}
(hereafter L12).

These authors use a Gaussian mixture model (GMM) to identify groups of
pulsars in the $P$~--~$\dot P$ plane.  In this approach, it is assumed
that the density of objects inside a group (a cluster) can be fitted
with a two-dimensional normal distribution (in linear or logarithmic
scale).  Here, we develop further the approach used by \nocite{lee}
L12\footnote{We thank Kejia Lee for providing the original code used
  in the study L12}.  We analyse the distributions in the space of
periods and period derivatives, both on logarithmic scale. In other
words, the sources belonging to every cluster have log-gaussian
distribution.  In the plots below, these groups of sources are
represented by $2$-$\sigma$ contours, and we call them ellipses, which
actually stands for ``a group selected by the GMM with log-gaussian
distribution of density of objects in both coordinates''.

We are interested in understanding the origin of the clusters found by
L12 for normal radio pulsars, as well as in checking if the GMM can be
used to differentiate between different population synthesis
calculations.  We present the analysis of several toy models with
well-understood properties, provide an additional analysis of the
observed population of the radio pulsars, and finally, apply the
cluster analysis to study the realistic synthetic sets of the data
produced by the advanced population synthesis models.

The paper is organized as follows. In the next section we briefly
describe the GMM method. In Section 3, the GMM method is applied to
different pulsar populations --- both, synthesized and observed. Our
population synthesis models and calculations are also presented in the
third Section. In Section 4, we discuss the effectiveness of the GMM
method to falsify theoretical population synthesis models and present
two examples of the application of this method to other sets of
sources.  Finally, we present our conclusions.

\section{Gaussian Mixture Model}

The GMM is an iterative method which allows one to find clusters of
objects that follow the Gaussian distribution. Each set of the
clusters describes the data with some probability. We are interested
in finding the one which gives a higher probability for a smaller
number of clusters.

 The GMM as an unsupervised learning algorithm has demonstrated
  its efficiency in different areas where it used to extract empirical
  knowledge from data samples \citep{gmm_eff1,gmm_eff2}.  Note that
  Gaussians on which the method is based are not basis functions.
  Nevertheless, the central limit theorem may be used to explain our
  motivation to use this method.  Indeed, a position of a pulsar in
  the $P$~--~$\dot P$ plane is a two-dimensional independent random
  variable because it is determined by the effective magnetic field
  and the age. The logarithm of the effective magnetic field is in the
  range $[10, 13]$, and the logarithm of the age is in the range $[3,
    10]$\footnote{Here it is reasonable to consider logarithms of
    quantities because we apply the GMM in the logarithmic
    scale.}. This defines a rather narrow variance range.  The central
  limit theorem states that a set of the large number of independent
  random variables where every variable belongs to a distribution with
  similar mean and variance has limiting distribution function
  approaching the normal distribution.  This is our main motivation to
  use Gaussians to study the pulsar distribution.

All calculations start with only one cluster. We then probe
sequentially $\sim$ 20 different initial positions of a cluster which
cover the entire data range.  It is found that the size of the initial
cluster does not change the results, while the position of the initial
cluster is crucial.   At the next step, we maximize the likelihood
  by varying the model parameters. The details of the procedure of
  likelihood maximization, which is called below the
  Expectation-Maximization algorithm \citep{press2007} are described
  in detail in L12. This procedure is run for all initial guesses for
  one cluster.  As a result, we have a few (typically 4-5 out of 20)
  sets of cluster parameters with significantly different
  properties.  For each of them we generate 10 samples of data
points, and for each such sample we carry out the two-dimensional
Kolmogorov-Smirnov (K-S) test.  The most probable set of cluster
parameters is used later as a fixed initial guess for the first
cluster.  Now, we add the second cluster.  For the second cluster, we
also make 20 initial guesses. The Expectation-Maximization algorithm
is run, and finally, after application of the K-S test, we fix the
initial guess for both clusters (note, that the initial guess for the
first cluster could change when we analyse a two-cluster model).
Following this algorithm we find all sets of cluster parameters
sequentially from one cluster to some maximum number of clusters.

Up to this moment we do not exclude any significantly distinct sets of
clusters.  When all final sets of clusters (in our study up to 4-5
clusters in each set) are found, we again generate 20 samples of data
points.

 In our numerical experiments, we found that if we use sets with
  more than 5 clusters then we obtain a large number of different
  solutions with equally high probabilities (this is called
  overfitting in L12).  On the other hand, when we limit the number of
  clusters to 4-5, we typically obtain no more than one or two
  solutions with comparably high probabilities.

We compare these samples to each other by means of the K-S test, and
then study the distribution of $D$'s (this is the Kolmogorov-Smirnov
statistic).

Finally, we calculate the probability that the data are drawn from the
synthetic sample which corresponds to the chosen set of clusters, and
the analysed data (observed or obtained by a population synthesis
model) are drawn from the same parent distribution based on the mean
value of $D$ for the analyzed data.

 As the final result we select the set of clusters which has the
  highest probability (higher than $\sim$90 per cent) and which
  contains the smallest number of clusters.  As it is mentioned in
  L12, the cumulative distribution for the two-dimensional K-S test is
  not well defined. Due to this, it was suggested to compute a
  cumulative distribution of $D$ for every model, and then we rely on
  this numerical statistics when we calculate the probability.  Based
  on our experience such numerical cumulative distribution does not
  provide better resolution than about a few percent. We say that the
  probability is 99.9 per cent when $D$ lies to the left from the
  begining (the first data point) of the cumulative distribution,
  i.e. in the region where our simulation does not have enough
  resolution.
 
\section{GMM for different populations}
\label{analys}

In this section we present the cluster analysis using the Gaussian
mixture model for several synthetic and observed populations of normal
radio pulsars.

First, we study the same population as in L12\nocite{lee} to be
certain that we can reproduce their results. Once we have made sure
that this is indeed the case, we apply the method to other
populations: the observed and the synthetic one.

In all cases, for a given set of the data points in the $P$~---~$\dot
P$ plane (observed or calculated) several runs of the
Expectation-Maximization algorithm are made for different sets of
initial guesses (see details in L12\nocite{lee}). These runs produce
different sets of the Gaussian ellipses (the number of ellipses can
also be different) which describe the data with some probability
defined according to the multi-dimensional Kolmogorov-Smirnov
test. Typically, we try to use as small a number of ellipses as
possible to describe the data with a reasonable probability, and we
demonstrate only one plot with the largest probability for the given
number of ellipses.

When a distribution of pulsars is analysed, the GMM method provides
multiple solutions, and to choose the most appropriate one it is
necessary to use some formal criteria. We use the K-S test probability
as such formal parameter.  Some sets of ellipses which do not have as
high probability as the presented ones, potentially can reflect
physically linked groups of pulsars better.  However, since {\it a
  priori} we do not have enough information, it is possible that we
can miss such good solutions.

\subsection{Analysis of observed populations}
\label{observed}

\begin{figure*}
\begin{minipage}{0.49\linewidth}
\includegraphics[width=84mm]{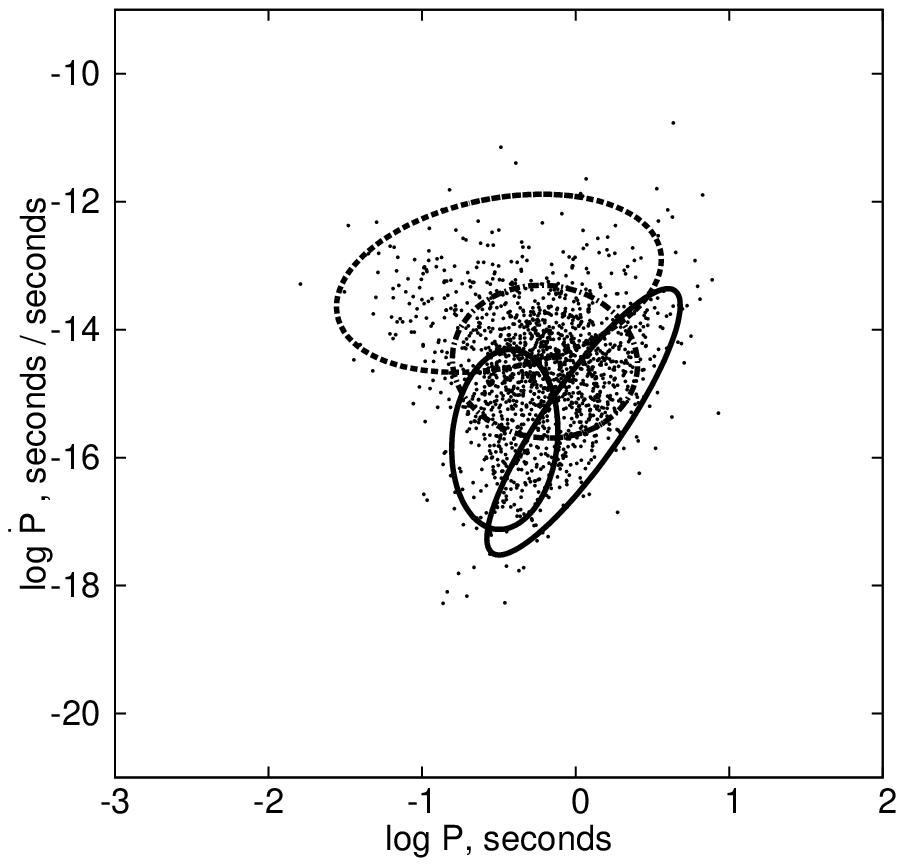}
\end{minipage}
\hfill
\begin{minipage}{0.49\linewidth}
\includegraphics[width=84mm]{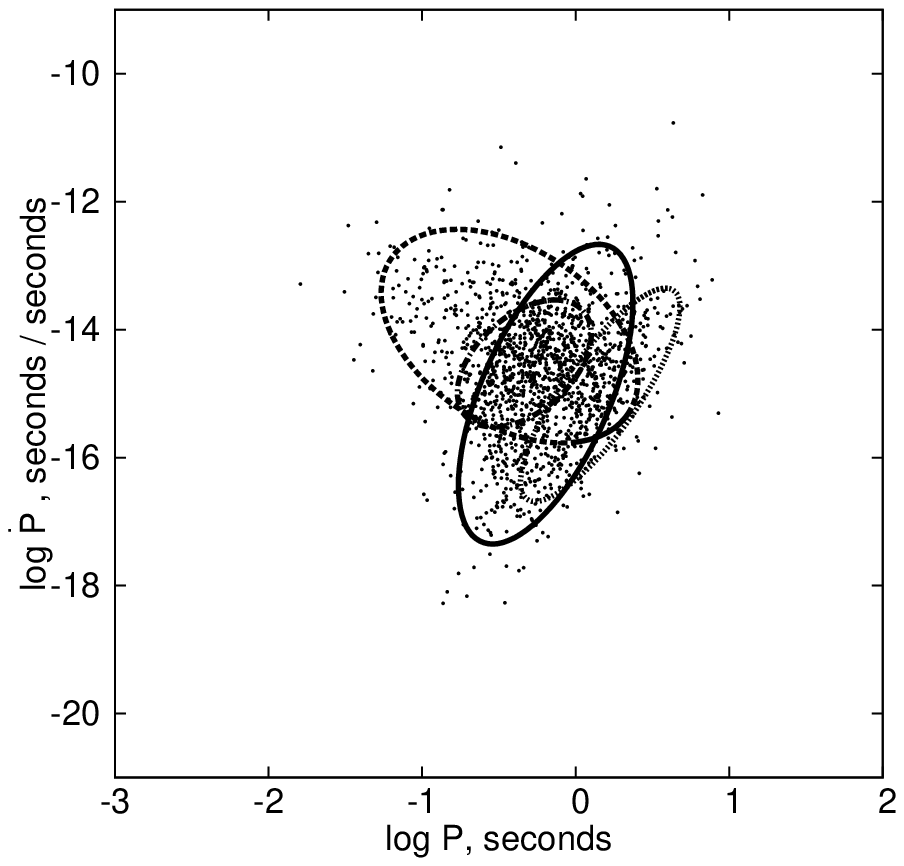}
\end{minipage}
\caption{$P$~--~$\dot P$ diagram for normal pulsars with ellipses
  derived with the GMM.  No magnetars, rotating radio transients,
  cooling near-by NSs, etc. are included into the data set.  The
  probability that the model and the data are drawn from the same
  parent distribution is 98 per cent for the model in the left panel,
  and 95 per cent for the model in the right panel.}
\label{norm_puls}
\end{figure*}

L12\nocite{lee} made their analysis for a mixed population where, in
addition to normal radio pulsars, sources like magnetars and close-by
cooling radio quiet neutron stars (NSs) were also considered.  We
exclude those objects which have been initially identified not as
radio pulsars, but as sources of a different nature (magnetars,
rotating radio transients, cooling near-by NSs, etc.).  Results are
shown in Figure~\ref{norm_puls}.  We present two most probable
realisations based on the same data set.

As in L12, \nocite{lee} four clusters are necessary to describe the
data.  First, let us note that both pictures are different from the
one in L12\nocite{lee}.  As we removed from the data most of the
sources with large $P$ and $\dot P$ (upper right part of the diagram:
all magnetars and close-by cooling NSs) the ellipse corresponding to
this part of the diagram in L12 is changed. However, this is not the
only modification. This is especially obvious in the left panel of
Figure~ \ref{norm_puls}, which has a higher probability: in this plot
one of the ellipses is shifted towards the region with smaller $\dot
P$ with respect to the figure in L12.  Only the ellipse lying along
the death line saves its position. In our opinion, this argues against
the effectiveness of the method to identify physically or evolutionary
related groups of normal radio pulsars.

\begin{figure*}
\begin{minipage}{0.49\linewidth}
\includegraphics[width=84mm]{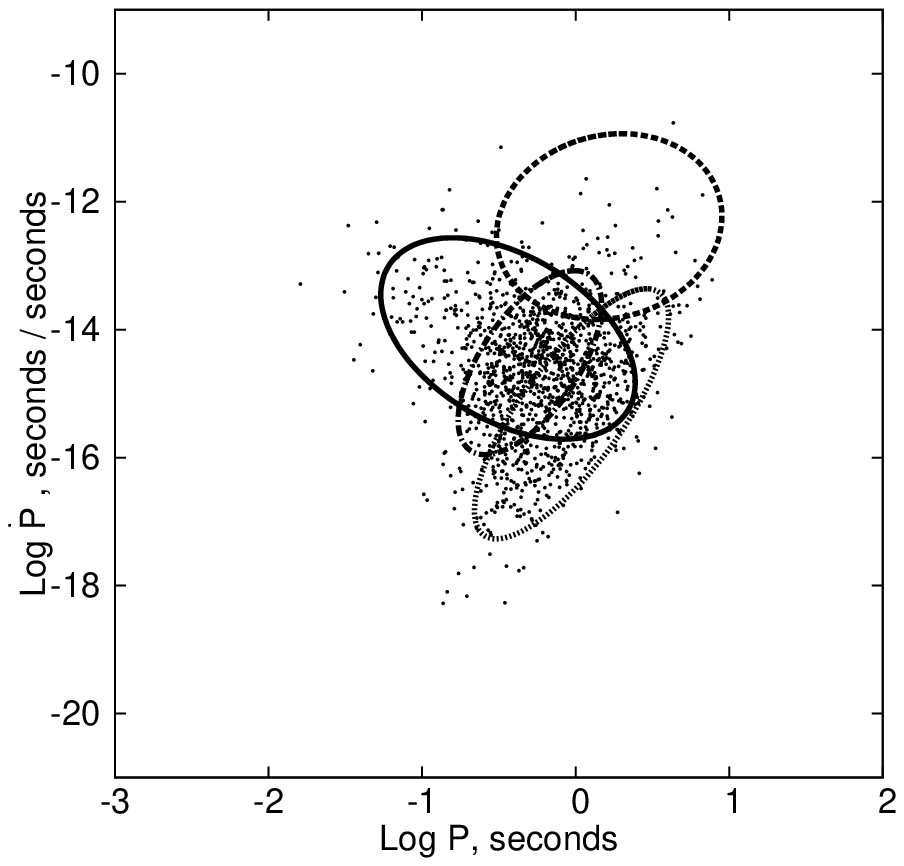}
\end{minipage}
\hfill
\begin{minipage}{0.49\linewidth}
\includegraphics[width=84mm]{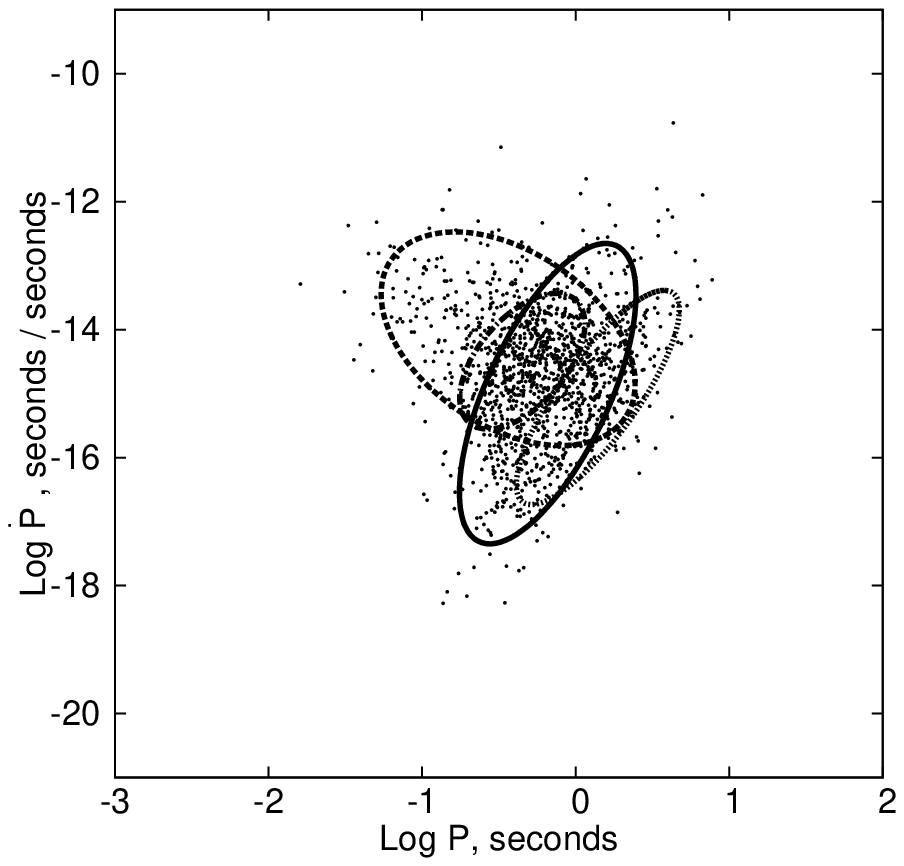}
\end{minipage}
\caption{The same as Figure 1, but 10 per cent of the objects are
  randomly removed.  The probability that the model and the data are
  drawn from the same parent distribution is 94 per cent for the model
  in the left panel, and 95 per cent for the model in the right panel.
}
\label{norm_puls_10}
\end{figure*}

We also tried to check what happens if the set of the pulsars
considered is slightly modified.  We randomly excluded 10 per cent of
the sources (Figure~\ref{norm_puls_10}).  Again, we see that the
results are changed.  In one (left) plot we obtained close to a
carbon-copy of the distribution from L12\nocite{lee}, though magnetars
are excluded here. In another, which has a very similar probability,
high-B pulsars do not form a separate group (note, that the data are
the same, only the initial combination of the ellipses is different).

\begin{figure}
\includegraphics[width=84mm]{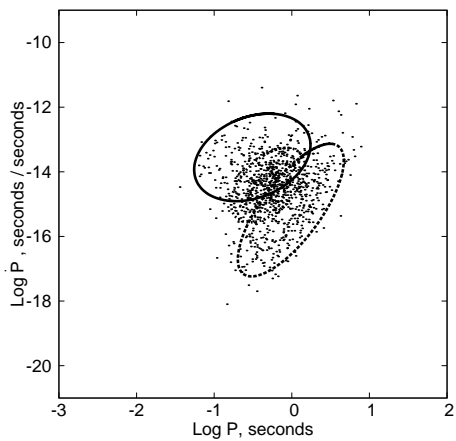}
\caption{The Parkes multibeam survey pulsars and the related clusters
  identified with the GMM. The number of data points is 905.  The
  probability that the model and the data are drawn from the same
  parent distribution is 94 per cent.  }
\label{parkes}
\end{figure}

Finally, we take only 905 pulsars detected in the Parkes multibeam
survey \citep{parkes2001} to get rid of some of the selection effects
(Figure~\ref{parkes}).  The picture is now simplified and becomes more
stable. One three ellipses are required. One is most probably related
to the existence of the death line. To see if we can find
interpretations for the others we run several population synthesis
models with different assumptions.

\subsection{Toy models of the pulsar evolution}
\label{toy}

One of the main tasks of this study is to try to understand the origin
of the clusters found for the observed populations. To do this in a
systematic manner we perform several runs for different synthetic
populations based on well-understood assumptions with regards to their
evolution, the spatial distribution, etc.  We use population synthesis
models of different complexity to see if an addition of a new option
(for example, the field decay or the Galactic spiral structure)
produces a new cluster or even changes the whole picture.

Let us briefly describe the basic features of our population synthesis
models.  Typically, pulsars are born in four spiral arms (unless the
opposite is explicitly stated).  The spiral arms are parametrized
according to \cite{vallee}.  The initial spin periods are taken from
the Gaussian distribution with $\langle P_0 \rangle = 0.3$ s and
$\sigma_p = 0.15$ s.  The initial magnetic fields are described by the
Gaussian distribution in log-scale: $\langle \log B_0/[\mathrm{G}]
\rangle = 12.65$ and $\sigma_B = 0.55$ (if opposite is not explicitly
stated). Pulsars are born with constant rate, the oldest have ages of
$3\, 10^8$ years.

The motion of every pulsar in the Galaxy is determined by its birth
place, the kick velocity, and the Galactic gravitational potential.
In all models we use the potential proposed by \cite{kuijkengilmore}.
The kick velocity distribution is chosen in the form: 
\be
p(v_l)=\frac{1}{2\langle v_l\rangle}\exp{\left[-\frac{|v_l|}{\langle
      v_l\rangle}\right]}.
\label{vel}
\ee 

Here $\langle v_l \rangle = 180$ $\mathrm{km\ s^{-1}}$
\citep{faucherkaspi2006}; $|v_l|$ denotes the absolute value of $v_l$.
Every component of the velocity vector is determined according to the
probability defined by Equation~(\ref{vel}). We neglect the Shklovskii
effect as well as the changes of the relative position of the Sun and
the spiral arms.

A pulsar is ``detected'' if its luminosity exceeds the limit
$S_\mathrm{min}$ for the Parkes multibeam survey
\citep{parkes2001}. The radio luminosity is calculated as
\citep{faucherkaspi2006}: \be \log L = \log (L_0 P^{\epsilon_P} \dot
P^{\epsilon_{\dot P}}_{15}) + L_\mathrm{corr}, \ee with $L_0 = 0.18
\ \mathrm{mJy}\ \mathrm{kpc}^2 $, $\epsilon _P = -1.5$, $\epsilon
_{\dot P}=0.5$. $L_\mathrm{corr}$ is a normally distributed random
function with a zero average and
$\sigma_{\mathrm{L}_\mathrm{corr}}=0.8$. $P$ is the spin period, and
$\dot P_{15}$ is the period derivative in units $10^{-15}$~s~s$^{-1}$.
The beaming fraction is calculated as in \cite{faucherkaspi2006}.  A
pulsar is detectable only above the death-line
\citep{ruderman1975,rawley1986}: \be \frac{B}{P^2} > 0.12\,
10^{12}\ \mathrm{G\ s^{-2},} \ee where $B$ is the equatorial magnetic
field.

In all runs the number of objects used in the analysis is equal to
905, i.e. equal to the number of the data-points from the Parkes
survey (see Fig. 3).  The analysis based on the GMM appears to be
dependent on the number of pulsars considered.  Therefore, we decide
to use the same number of pulsars in all generated samples as it is in
the Parkes multibeam survey (see \ref{observed}).

All models are listed in the Table \ref{mod}, and short comments are given for
each.

\begin{table}
\caption{List of synthetic models. 
}
\label{mod}
\begin{tabular}{@{}lclrl}
\hline
Name & Fig.

        & Electron density
	
	& Prob.

	& Remarks   \\

\hline

Model I.    & \ref{nd_A1}       & NE2001   & 0.99  & --  \\					%  nd\_A1
Model II.   & \ref{A0626}       & $DM=15D$ & 0.89  & --  \\				%  A0626
Model III   & \ref{d_A1}        & NE2001   & 0.999 & exp. decay (Eq. \ref{dec}) \\ 			%  d\_A1\_905  
Model IV.   & \ref{nd_A1_nosel} & $DM=15D$ & 0.96  & no spiral arms  \\						% nd\_A1\_nosel\_15
Model V.    & \ref{A1137}       & $DM=15D$ & 0.92  & increased $R_\mathrm{Gal}$\\%A1137
Model VI.   & \ref{D3}          & NE2001   & 0.94  & bimodal in $P_0$  \\		% D3
Model VII.  & \ref{D5}          & NE2001   & 0.97  & bimodal in $B_0$  \\			% D5 
Model VIII. & \ref{popov}       & NE2001   & 0.98  & \cite{popovpons},  \\					% Popovnd\_dl
            &                   &          &       & no decay\\
Model IX.   & \ref{popov}       & NE2001   & 0.96  & \cite{popovpons},   \\ 			% Popovd\_dl
            &                   &          &       & decay \\
\hline

\end{tabular}
\end{table}

\subsubsection{The simplest toy models}
\label{simplest_toy}

We start with the simplest models with the standard magneto-dipole
losses with a constant angle between the spin and the magnetic dipole
axis. The magnetic field is constant, too.

\begin{figure}
\includegraphics[width=84mm]{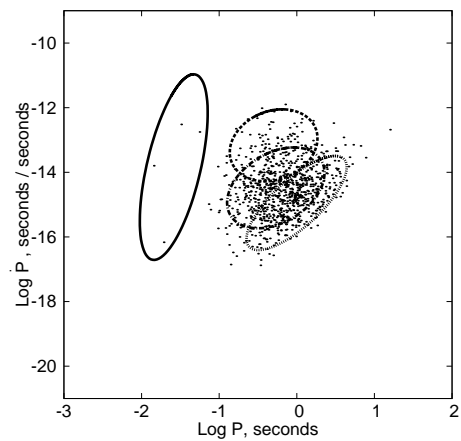}
\caption{Model I. The magnetic field is constant. Pulsars are born in
  spiral arms. The electron density is calculated with the NE2001
  model.  The probability that the model and the data are drawn from
  the same parent distribution is 99 per cent.}
\label{nd_A1}
\end{figure}

In the first model that we present, the electron density is calculated
according to the NE2001 \citep{cordeslazio}. The data can be well
described with four clusters, see Figure~\ref{nd_A1}.  Three clusters
are rather similar to those we see in the fit for the Parkes multibeam
survey data (Figure~\ref{parkes}).  The left cluster, containing just
four pulsars, breaks this similarity.  Note that this cluster is
statistically significant. Indeed, the GMM finds no similar models
with just three clusters. And the most probable model with three
clusters is drawn from the same parent distribution as the data with
the probability 92 per cent (compare with 99 per cent for 4 clusters).
 
The results presented in Figure~\ref{nd_A1} reproduce all main
features seen in the plot for the Parkes data (Figure~\ref{parkes}).
However, slightly different runs of the population synthesis code,
where the integration of the individual pulsar histories lasted not
for $3\, 10^8$~yrs, but less, for example $2\, 10^8$~yrs, produce very
different results in terms of the identified clusters.  In our
opinion, this demonstrates that even for the simple models with well
understood ingredients the method does not produce a stable set of
easily interpretable clusters.

We can simplify the Model I by using a less complex model for the
electron density distribution in the Galaxy.  In the Model II, the
dispersion measure is calculated with a simple relation: \be DM=15 \,
D.
\label{dm_calc}
\ee

Here $D$ is the distance in kpc. The minimum brightness is dependent
on the interstellar scattering time, $\tau_\mathrm{scat}$.  If the
$DM$ is taken in the form (\ref{dm_calc}), then $\tau_\mathrm{scat}$
is calculated according to Equation (7) from \cite{scatt}.

\begin{figure}
\includegraphics[width=84mm]{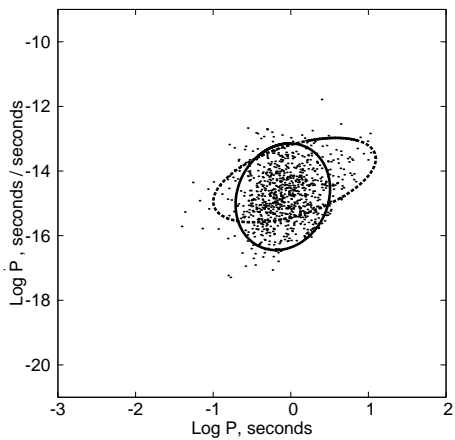}
\caption{Model II. The same as Model I in Fig.\ref{nd_A1}, but instead
  of the NE2001 model we use a simple linear relation between distance
  and dispersion measure (Eq. \ref{dm_calc}).  The probability that
  the model and the data are drawn from the same parent distribution
  is 89 per cent.  }
\label{A0626}
\end{figure}

If we compare Models I and II, the results are changed significantly.
In the simpler model, only two clusters are necessary to describe the
data, see Figure~\ref{A0626}.  We conclude that the effects connected
with the fluctuations of the electron density are strong, and
influence the number and the distribution of ellipses quite
significantly.

In addition, we made calculations for the modified simplest Model IV
when the electron density distribution is calculated according to
NE2001.  We do not present the figure, but the results are changed
significantly.  This again confirms that the influence of electron
density distribution on the picture of gaussian clusters is essential.

%It is possible to add realistic electron density distribution  
%(i.e. the NE2001 model)
%to our simplest Model~IV. We perform calculations, but do not
%include the final figure in the article because of its insignificance.
% Let us briefly describe the result. This model contains three clusters. 
%One is for main group of pulsars and other are located in upper 
%part of $P$~--~$\dot P$. This disposition of ellipses differs 
%from all other our models significantly.  

In principle, the identification of just two clusters in a set of the
observational data can lead to a conclusion of some dichotomy either
in the initial properties or in the evolutionary laws.  Here,
obviously, the pulsars are born from a single mode population. The
evolution also does not contain any process which can result in a
dichotomy of sources. Therefore, the origin of such dichotomy is
puzzling. We will simplify the model even more in Section~
\ref{more_simpl} in an attempt to clarify it.

\subsubsection{Field decay}

\begin{figure}
\includegraphics[width=84mm]{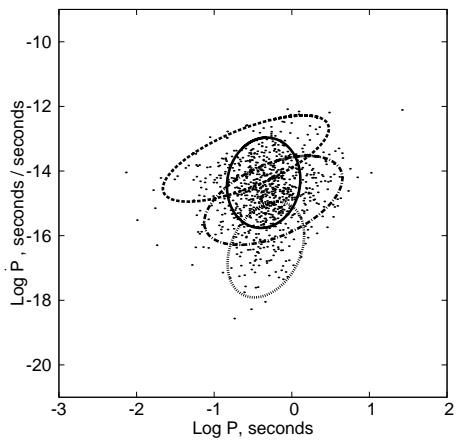}
\caption{Model III. Similar to Model I, but the magnetic field decays
  exponentially.  The probability that the model and the data are
  drawn from the same parent distribution is 99.9 per cent.  }
\label{d_A1}
\end{figure}

Now we add to our scenario the magnetic field decay. The first model
of this type (Model III) is similar to the one shown in
Figure~\ref{nd_A1}, but the field decays exponentially.  For the
illustrative purposes we choose a simple model:
\begin{equation}
B(t) = B_0 \, \exp\left[{-\frac{t}{\tau_\mathrm{mag}}}\right].
\label{dec}
\end{equation}
Here $\tau_\mathrm{mag} = 5\, 10^6$ years.  We see
(Figure~\ref{d_A1}), that the set of ellipses is very different from
the Model I.  In most cases the pulsars with larger $\dot P$ are
joined in one Gaussian with the pulsars from the main population, so
it is difficult to identify if there is a magnetic field decay in this
model, looking at the distribution of the ellipses.

\subsubsection{Spatial distribution}
\label{more_simpl}
\begin{figure}
\includegraphics[width=84mm]{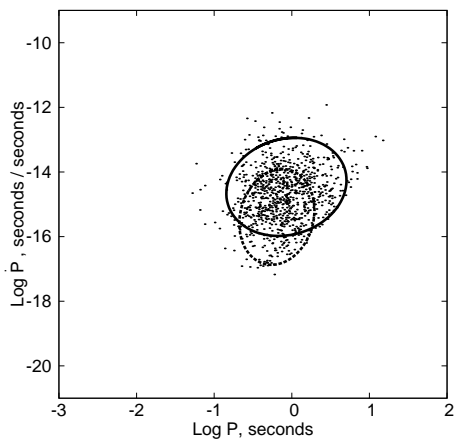}
\caption{Model IV. Similar to Model II, but the birth places of the
  pulsars are not related to the spiral arms.  The probability that
  the model and the data are drawn from the same parent distribution
  is 96 per cent.  }
\label{nd_A1_nosel}
\end{figure}

Let us make further simplifications of the Model II presented in
Figure~\ref{A0626}. Now we exclude the spiral structure. This is the
simplest model we study here (Model IV), see Fig. \ref{nd_A1_nosel}.
The picture looks only slightly different.

\begin{figure}
\includegraphics[width=84mm]{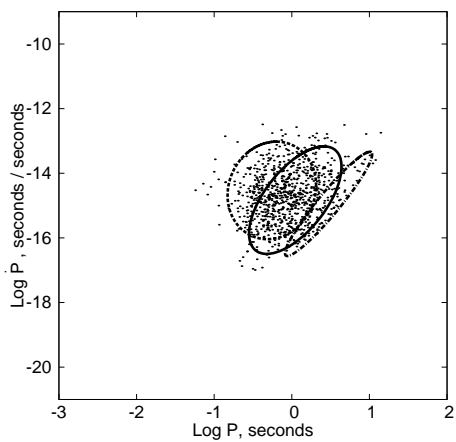}
\caption{Model V. Similar to Model II, but the size of the Galaxy and
  the total birth rate are increased.  The probability that the model
  and the data are drawn from the same parent distribution is 92 per
  cent.  }
\label{A1137}
\end{figure}

Another modification related to the spatial distribution is very
artificial. We want to study how the finite size of the Galaxy
influences the picture. Therefore, we increase the size of the Galaxy,
$R_\mathrm{Gal}$, by a factor of three, and increase the pulsar
production rate respectively.  The results for the Model V are
presented in Figure~\ref{A1137}.  Now the distribution is described
with three clusters and two of them are elongated along the
death-line.

\subsubsection{Initial distributions}

In this subsection we present two models with modified initial
distributions of the spin periods and the magnetic fields. Instead of
using a single Gaussian for each of these parameters, in one case we
use two well-separated Gaussian distributions for the initial spin
periods; in another we use two Gaussians for the magnetic fields
(which are assumed to be constant during evolution).

In our Model VI the initial spin period distribution consists of two Gaussians:
$\langle P_0\rangle =0.02$~s,
$\sigma_\mathrm{P}=0.02$~s, and
$\langle P_0\rangle =0.2$~s,   
$\sigma_\mathrm{P}=0.05$~s.

\begin{figure}
\includegraphics[width=84mm]{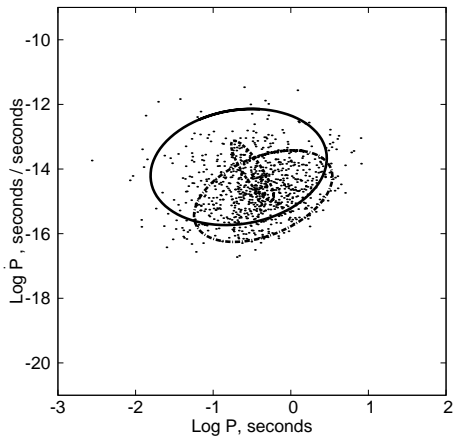}
\caption{Model VI. The initial spin period distribution consists of
  two Gaussians: $\langle P_0\rangle =0.02$~s,
  $\sigma_\mathrm{P}=0.02$~s, and $\langle P_0\rangle =0.2$~s,
  $\sigma_\mathrm{P}=0.05$~s.  The probability that the model and the
  data are drawn from the same parent distribution is 94 per cent.  }
\label{D3}
\end{figure}

Three clusters are identified, see Figure~\ref{D3}.  They overlap
sufficiently. We suppose that for the GMM it is difficult to
distinguish pulsars from different sub-populations because of their
significant mixing. After a short time (about $ 10^4$~--~$10^5$ years)
pulsars with shorter periods are braked enough and can be confused
with younger pulsars from the second sub-population with longer
initial periods.

\begin{figure}
\includegraphics[width=84mm]{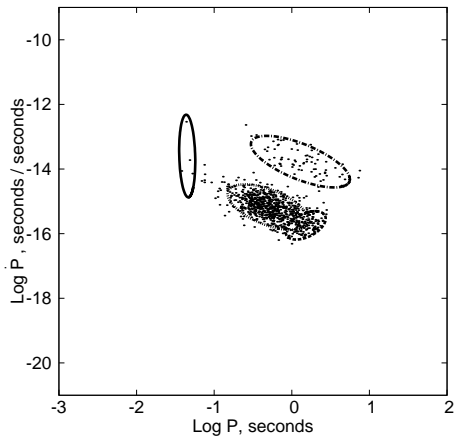}
\caption{Model VII.  The initial magnetic field distribution is
  bimodal: $\log B_0/[\mathrm{G}]=11.8$, $\sigma_\mathrm{B}= 0.15$ and
  $\log B_0/[\mathrm{G}] = 12.7$, $\sigma_\mathrm{B}=0.15$.  The
  probability that the model and the data are drawn from the same
  parent distribution is 97 per cent.  }
\label{D5}
\end{figure}

The picture of the clusters distribution for the case of the bimodal
distribution of the initial magnetic field could look simpler as
pulsars with stronger magnetic fields are higher in the $P$~--~$\dot
P$ plane, and they are never mixed with low-magnetized pulsars without
the field decay.  Nevertheless, in the case of two Gaussians for the
initial magnetic field we have to make them very well separated to
identify these groups using the GMM-code. When the first Gaussian is
centred on $\log B_0/[\mathrm{G}]=12.65$ and the second --- on $\log
B_0/[\mathrm{G}]=13.2$ (both with $\sigma_\mathrm{B}= 0.55$), the GMM
is not able to distinguish them.  Even for $\log
B_0/[\mathrm{G}]=12.3$ and $\log B_0/[\mathrm{G}]=13.2$ (both with
$\sigma_\mathrm{B}=0.35$) the method does not work well.  Only when we
take $\log B_0/[\mathrm{G}]=11.8$ and $\log B_0/[\mathrm{G}]=12.7$
with $\sigma_\mathrm{B}=0.15$ (Model VII) the two groups are clearly
described by different ellipses.  However, in this case it was also
easily visible by eye, see Figure~\ref{D5}.  Note, that the additional
clusters are also necessary to describe the data, similar to Model I.

\subsection{Analysis of the advanced synthetic models}
\label{realistic}

It is interesting to apply the GMM to the results of the advanced
models of the population synthesis. We use results of the calculations
using the code of \cite{popovpons}\footnote{We thank prof. J.A. Pons
  for providing these data sets.  Note, that the exact realisations
  used here are different from those presented in the paper by Popov
  et al.  (2010)\nocite{popovpons}.}.

\begin{figure*}
\begin{minipage}{0.49\linewidth}
\includegraphics[width=84mm]{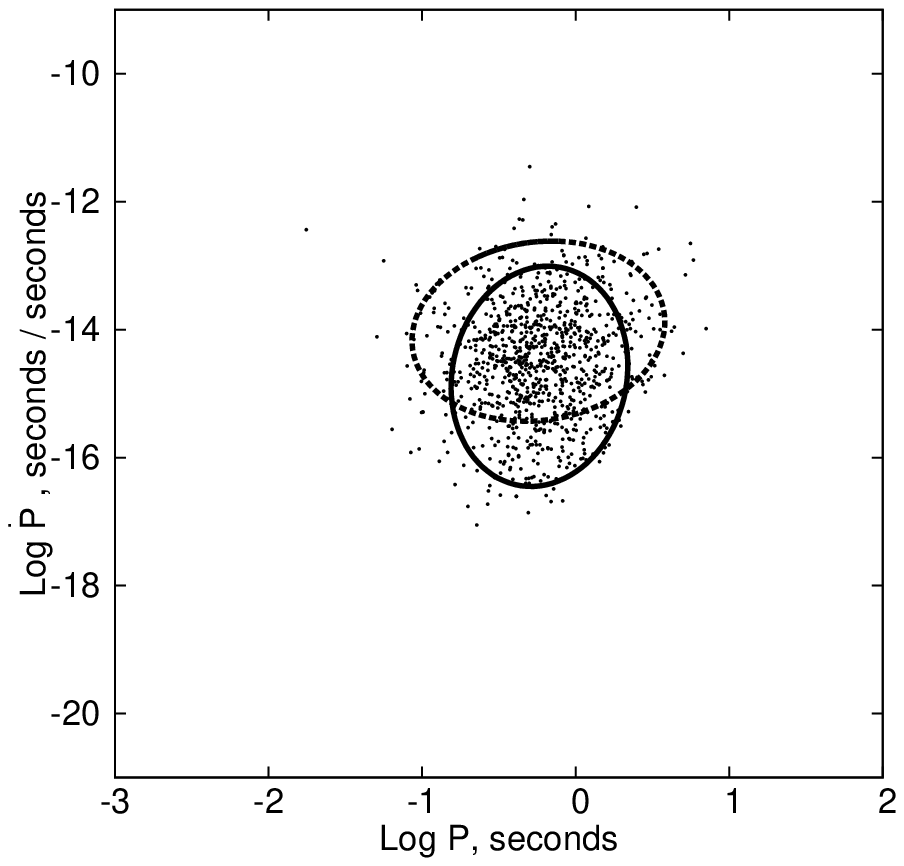}
\end{minipage}
\hfill
\begin{minipage}{0.49\linewidth}
\includegraphics[width=84mm]{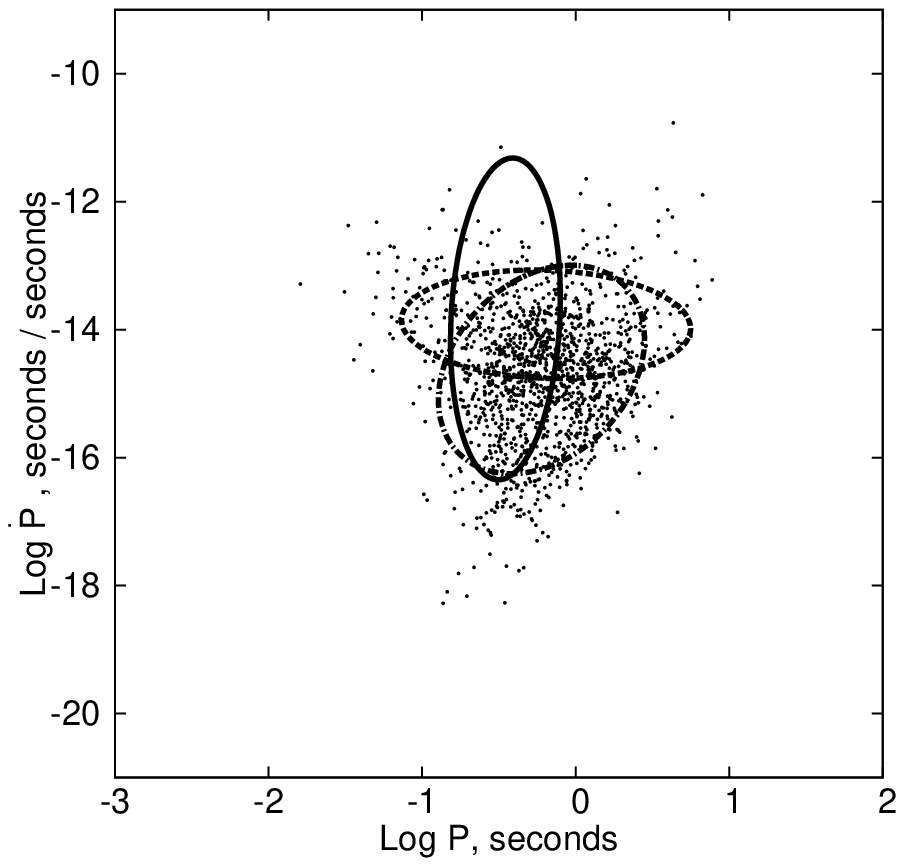}
\end{minipage}
\caption{Models VIII (left panel) and IX (right panel). The data was
  calculated with the population synthesis model used in Popov et
  al. (2010). In the left panel pulsars evolve with constant magnetic
  fields.  The K-S test probability for this model is 98 per cent.  In
  the right panel results for decaying fields are presented.  For this
  model the K-S test probability is 96 per cent.  }
\label{popov}
\end{figure*}

We study two data sets: one with and one without the magnetic field
decay.  Both are fitted to reproduce the properties of the observed
population.  The results are presented in Figure~\ref{popov}. In the
case of the constant field (Model VIII, left panel), the results can
be described with just two clusters, as in the case of the most simple
Model IV (Figure~\ref{nd_A1_nosel}), but the ellipses have different
properties. If the field is allowed to decay (Model IX), then we need
three clusters. However, the ellipses now are clearly determined by
the outlying points. The method does not distinguish the group of the
initially higher magnetized pulsars which experienced significant
field decay.  Note, that in the original calculations no death-line
was used, only the selection related to the flux, so no ellipse
stretched along the death-line is seen.  Note however, if we exclude
points behind the death line, then the pictures is not changed
significantly.

\section{Discussion}
\label{disc}

\subsection{How effective is the Expectation-Maximization algorithm for the radio
pulsars studies?}

Based on the significant amount of numerical experiments we can
conclude that the GMM method is not effective for the studies of
normal radio pulsars.  Let us try to speculate why this is the case.

The main initial distributions of pulsars in most of our models is
defined by two Gaussians: one for the spin periods and the second ---
for the magnetic fields (in log-scale). However, the GMM method is
applied to analyze the distributions in the $P$~--~$\dot P$ not in the
$B$~--~$P$ plane. For the standard magneto-dipole braking ($n=3$)
$\dot P$ can be expressed as:
\begin{equation}
\dot P = \frac{\alpha B^2}{P}.
\label{dotP}
\end{equation}
Here $\alpha=8\pi^2R^6/(3Ic^3)=10^{-39}\mathrm{\ cm\ s^3\ g^{-1}}$,
$R$ is the NS radius, $I$ is the moment of inertia, and $c$ is the
speed of light. Therefore, the value of $\dot P$ does not follow the
Gaussian distribution (in log-scale or not) from the very beginning of
our simulations. If the evolution does not follow the magneto-dipole
formula the argument is also true for the models we used.  In
particular, it explains why the GMM method fails to describe with a
single cluster the ensembles of pulsars when the maximum duration of
the evolutionary track followed was short. For example, we performed
the calculations for the maximum track duration equal to $\sim
2\ 10^7$ years. And in this case the results cannot be explained by a
single cluster.

Simple replacement of $\dot P$ by $B$ does not improve the
situation. Pulsars are rapidly evolving objects, and in the
$P~$--~$\dot P$ plane we can find objects at different stages of their
evolution. The tracks of pulsars in the $P$~--~$\dot P$ plane in the
simple case of the constant effective magnetic field are straight
lines. The length of the line is determined by the age, the initial
spin period and the initial magnetic field.  Therefore, the longer
lines are in the upper part of the $P$~--~$\dot P$ plot.  During the
evolution, the Gaussian distribution of pulsars in the $P$~--~$B$
plane shifts and rotates.  In addition to the already evolved pulsars,
there are new ones being born constantly.  Consequently, the
distribution would not follow a Gaussian distribution even in the case
of $P$~--~$B$ plot.

L12 \nocite{lee} discussed the problem of robustness of the method in
their study, too.  The authors show that if 3 per cent of all pulsars
are randomly removed from the dataset, no significant changes appear
in the structure of the cluster distribution.  In our opinion, 3 per
cent is a small number.  New observations routinely bring many tens,
or even hundreds of newly discovered objects.  So, 10 per cent is a
more realistic number to study the stability of the method.  It is
shown above that a 10 per cent modification of the number of pulsars
changes the results significantly. In addition, as we have
demonstrated, a systematic exclusion of even a small numer of objects
(magnetars etc. in the studied case) also changes the set of the
ellipses significantly.  This brings us to the conclusion that the
method is not very effective.
% even though 
%the number of pulsars in  cluster F in analysis by 
%L12 \nocite{lee} is about 3 per cent of all objects.
%Ordinarily a new survey supplys much more 
%than 10 per cent of newly detected pulsars.
%Moreover, it 10 per cent is the standard 
%fluctuation in full number of pulsars which
%gives the population synthesis method. 

\subsection{Other examples}

In our opinion, the GMM is not very effective in distinguishing
physically or evolutionary related groups when applied to rapidly
evolving populations observed with significant selection effects. That
is why we decided to apply the method to more ``stable'' sets of
data.\footnote{In L12 it was demonstrated that the method can be
  successfully used for the millisecond pulsars, which evolve much
  more slowly compared to the normal radio pulsars.}

First, we decided to use the GMM-code to study the well-known
distribution of the gamma-ray bursts (GRBs) in the duration-hardness
diagram. It is well established that there are at least two
populations: the long soft and the short hard bursts \citep{kleb92,
  kouv93}.  The duration and hardness of GRBs data are obtained from
the BATSE 4B Gamma-Ray Bursts Catalog \citep{batse}. We applied the
GMM method to the distributions of bursts by logarithm of duration
($t_\mathrm{90}$) and by logarithm of hardness. The hardness is
defined as
\begin{equation}
S = \frac{F_\mathrm{100-300\ keV}}{F_\mathrm{50-100\ kev}},
\end{equation}
where $F_\mathrm{100-300\ keV}$ is the flux in the energy range
100-300 keV, and $ F_\mathrm{50-100\ kev}$ is the flux in the range
50-100 keV.

\begin{figure*}
\begin{minipage}{0.49\linewidth}
\includegraphics[width=84mm]{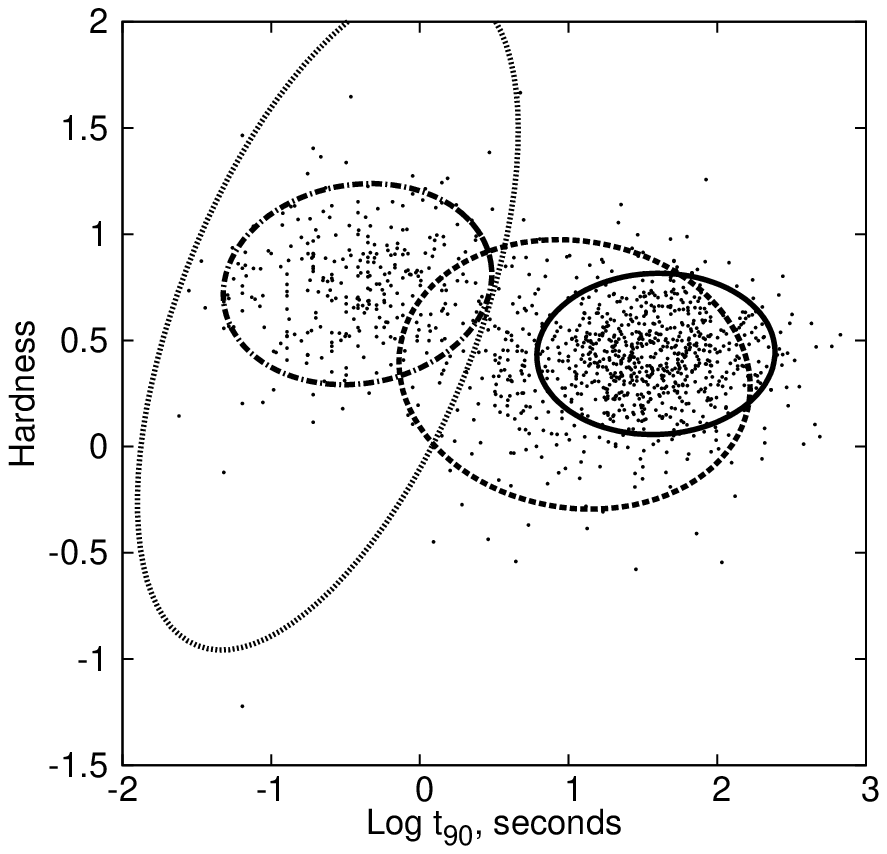}
\end{minipage}
\hfill
\begin{minipage}{0.49\linewidth}
\includegraphics[width=84mm]{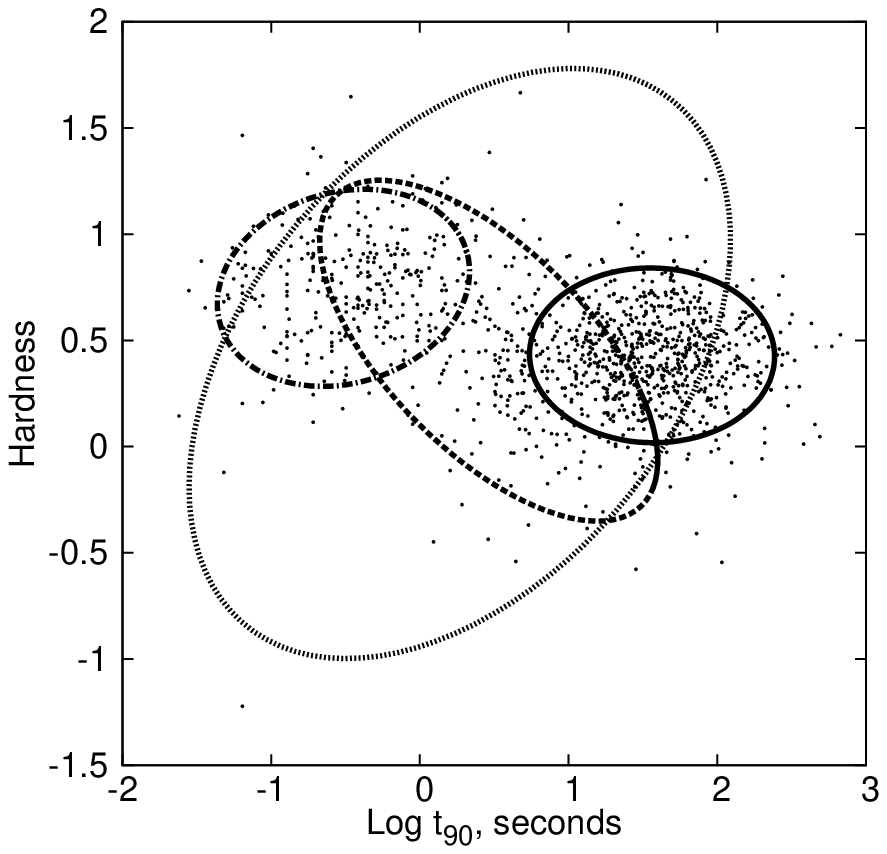}
\end{minipage}
\caption{Gamma-ray bursts analysis. Two realizations of the
  Expectation-Maximization method are shown. In both cases, the
  well-known bimodal distribution in the duration-hardness cannot be
  described by two Gaussian clusters.  The probability that the
  presented model and the data are drawn from the same parent
  distribution is 99 per cent for the model in the left panel, and 97
  per cent for the model in the right panel.  }
\label{batse}
\end{figure*}

Naively, one expects that the method will easily describe the
distribution with two clusters which correspond to the two GRB
types. But this is not the case! In Figure~\ref{batse} we see that the
method requires four clusters.  Depending on the initial guess, a
different configuration of ellipses with nearly equal likelihood can
appear, and in some of them the ellipses overlap.

For another test we choose the high-mass X-ray binaries, in particular
the Be/X-ray systems.  For these, the existence of two types have been
established by \cite{knigge2011}.  We want to check, whether the
Expectation-Maximization algorithm can also identify this
dichotomy. The data on Be/X-ray binary systems is taken from the
catalog by \cite{xray}\footnote{The catalog is available on-line at
  http://xray.sai.msu.ru/~raguzova/BeXcat/}.

\begin{figure*}
\begin{minipage}{0.49\linewidth}
\includegraphics[width=84mm]{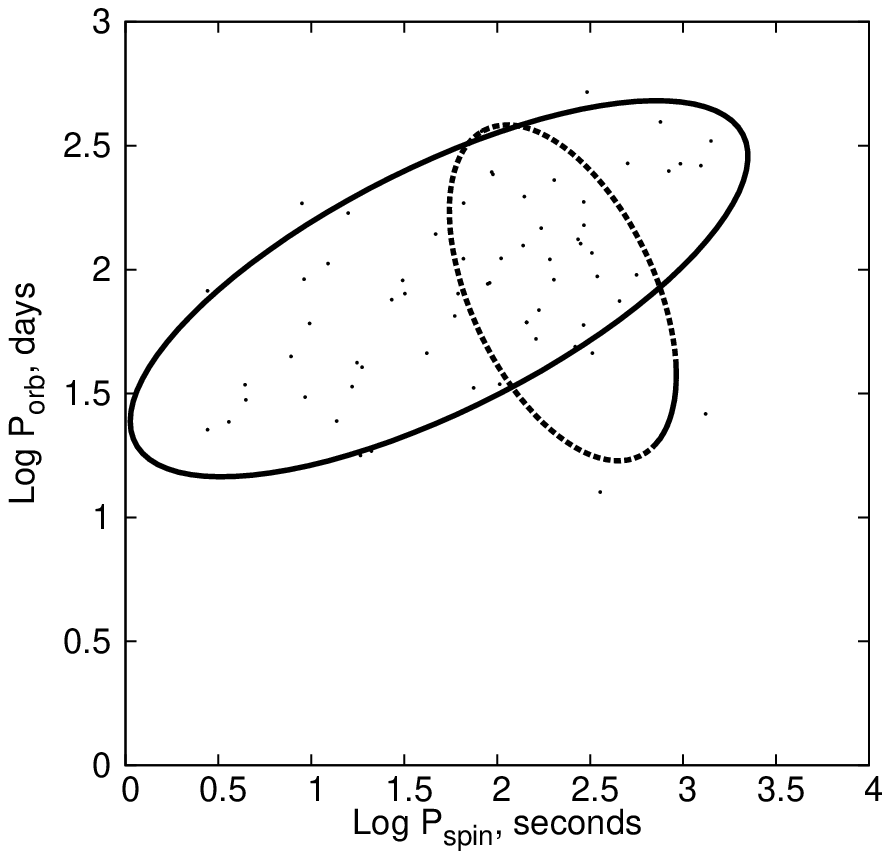}
\end{minipage}
\hfill
\begin{minipage}{0.49\linewidth}
\includegraphics[width=84mm]{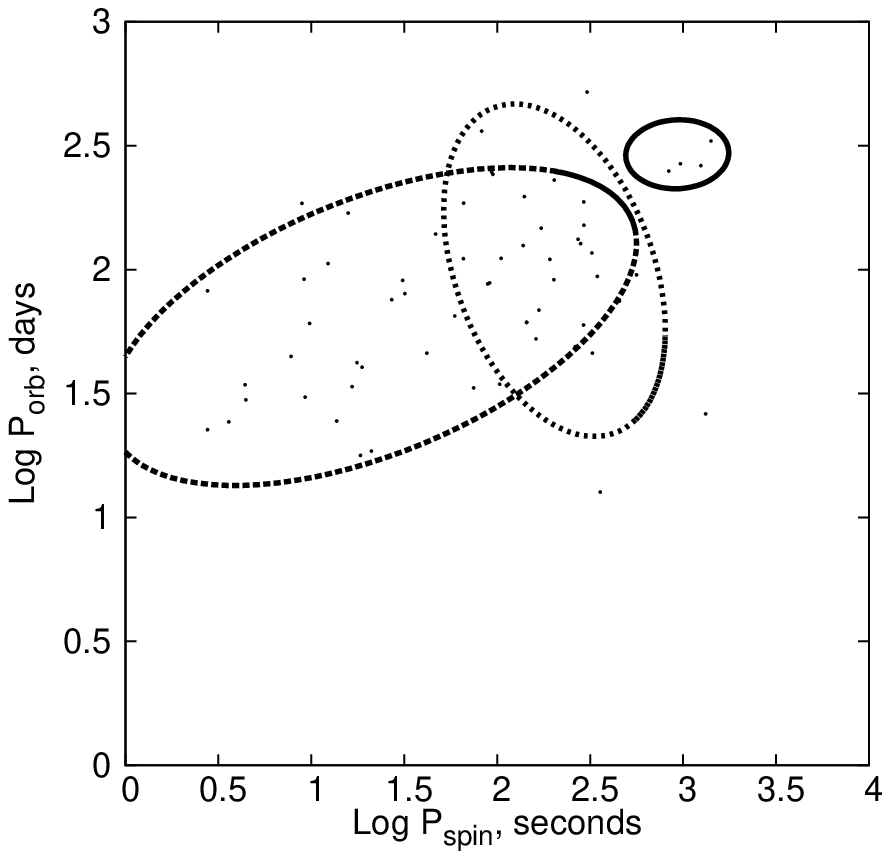}
\end{minipage}
\caption{Analysis of the $P_\mathrm{spin}$~--~$\dot P_\mathrm{orb}$
distribution 
of  Be/X-ray binaries with the GMM method.
The probability that the model and data are drawn from the same parent distribution
is 99.9 per cent for both  models.
% in the left panel,  and 99.9 per cent for
%the model in the right panel.
}
\label{xrb}
\end{figure*}

The GMM method was applied to the distribution of Be/X-ray pulsars by
logarithm of the spin period and the logarithm of the orbital period.
The results are shown in Figure~\ref{xrb}. Indeed, two Gaussians are
enough to describe the data. However, we do not see that the method
separates sources into long (both, the orbital and the spin) and the
short periods. Instead, the short period sources are united with those
with the longest periods. We have to note here, that the objects with
the longest periods that we use were not included in the sample in
\cite{knigge2011}.

\section{Conclusions}
\label{concl}
In our study, the GMM method was at first applied to the distributions
of the known normal radio pulsars in the $P$~--~$\dot P$ plane.  It is
found that the exclusion of magnetars, the thermally emitting NSs,
etc. modifies the set of the clusters as compared to those found by
L12\nocite{lee}. The pulsars detected in the Parkes multibeam survey
may be described with just three clusters. It was also found that the
results of the Gaussian cluster finding procedure are not robust (or
the GMM method is oversensitive and the small changes in the data
cause significant changes in the results).  The relative position of
the ellipses were changed when we excluded random 10 per cent of the
pulsars from the observational data set.

In the second part of our study we generated ensembles of pulsars
using population synthesis models of different complexity.  First, we
find that the GMM method is strictly dependent on the total number of
pulsars in the analyzed ensemble.
The choice of the electron density model also has a strong influence
on the cluster distribution.  The spiral structure of our Galaxy has a
smaller effect.  Such features as the bimodal distribution of the
initial parameters are hardly recognized by the GMM method for the
realistic choices of parameters.  The magnetic field decay changes the
distribution of the clusters. Typically, if the field is decaying it
is necessary to use more clusters to describe the data.

We conclude that the GMM is not effective to test models of the normal
radio pulsar evolution because of the method's over-sensitivity.

\section*{Acknowledgments}
We thank Kejia Lee for numerous useful comments during the whole work on
this study and on the text of the manuscript. Special thanks to Vasily
Belokurov, who carefully read the paper which helped to improve its style.
We also thank the referee for useful remarks and suggestions.
The work of S.P. was supported by the RFBR grant 12-02-00186.
The work of A.P. was supported by Saint Petersburg University grant
6.38.73.2011.

\bibliographystyle{mn2e} 
\bibliography{ip}

\end{document}